\pdfoutput=1

\documentclass[fleqn,usenatbib]{mnras}

\usepackage{newtxtext,newtxmath}

\usepackage[T1]{fontenc}
\usepackage{ae,aecompl}

\usepackage{graphicx}	
\usepackage{amsmath}	
\usepackage{amssymb}

\usepackage{xcolor}
\usepackage[normalem]{ulem}

\definecolor{darkgreen}{rgb}{0.13, 0.55, 0.13}
\definecolor{rubinered}{rgb}{0.82, 0.0, 0.34}
\definecolor{orange(ryb)}{rgb}{0.98, 0.6, 0.01}
\definecolor{scut}{rgb}{0.58, 0.22, 0.51}

\title[Star formation thresholds]{Determining Star Formation Thresholds from Observations}

\author[S. Khullar et al.]{
Shivan Khullar,$^{1,2}$\thanks{E-mail: shivankhullar@gmail.com (SK)}
Mark R. Krumholz,$^{1,3}$
Christoph Federrath$^{1,3}$ \newauthor
\ and Andrew J. Cunningham$^{4}$
\\
$^{1}$Research School of Astronomy and Astrophysics, Australian National University, Canberra, ACT~2611, Australia\\
$^{2}$Department of Physics, Goa Campus, Birla Institute of Technology and Science, Pilani, Rajasthan, 333031, India\\
$^{3}$ARC Centre of Excellence for All-Sky Astrophysics in Three Dimensions (ASTRO-3D), Australia\\
$^{4}$Lawrence Livermore National Laboratory, Livermore, CA 94550, USA}
\date{Accepted XXX. Received YYY; in original form ZZZ}

\pubyear{2019}

\begin{document}
\label{firstpage}
\pagerange{\pageref{firstpage}--\pageref{lastpage}}
\maketitle

\begin{abstract}

Most gas in giant molecular clouds is relatively low-density and forms star inefficiently, converting only a small fraction of its mass to stars per dynamical time. However, star formation models generally predict the existence of a threshold density above which the process is efficient and most mass collapses to stars on a dynamical timescale. A number of authors have proposed observational techniques to search for a threshold density above which star formation is efficient, but it is unclear which of these techniques, if any, are reliable. In this paper we use detailed simulations of turbulent, magnetised star-forming clouds, including stellar radiation and outflow feedback, to investigate whether it is possible to recover star formation thresholds using current observational techniques. Using mock observations of the simulations at realistic resolutions, we show that plots of projected star formation efficiency per free-fall time $\epsilon_{\rm ff}$ can detect the presence of a threshold, but that the resolutions typical of current dust emission or absorption surveys are insufficient to determine its value. In contrast, proposed alternative diagnostics based on a change in the slope of the gas surface density versus star formation rate surface density (Kennicutt-Schmidt relation) or on the correlation between young stellar object counts and gas mass as a function of density are ineffective at detecting thresholds even when they are present. The signatures in these diagnostics sometimes taken as indicative of a threshold in observations, which we generally reproduce in our mock observations, do not prove to correspond to real physical features in the 3D gas distribution.
\end{abstract}


\section{Introduction}
\label{sec:Intro}

Understanding the physical factors behind the formation of stars from interstellar gas is key to developing a predictive theory of star formation and understanding the evolution of galaxies. Star formation is known to occur in filamentary structures \citep{Goldsmith08filament,A14} in molecular clouds \citep{WongBlitz2002,Kennicutt2007,Blanc2009,Krumholz14} and it can be characterized by a quantity known as the star formation efficiency per free-fall time $\epsilon_{\rm ff}$ \citep{KM05}, which measures the fraction of the gas that is converted to stars per free-fall time.
On molecular cloud scales, the average value of $\epsilon_{\rm ff}$ is known to be small, $\approx$ 0.01\footnote{As discussed in the \citet{Krumholz18a} review, the amount of spread about this average is a subject of current debate; \citeauthor{Krumholz18a} argue that the weight of evidence favours a relatively small spread of $\approx 0.3$ dex, but some authors argue for larger spreads of $\gtrsim 1$ dex.} (e.g., \citealt{KrumholzTan2007,KDM12,F13,Evans14a,SFK15,Vutisalchavakul16a,Heyer16a,Leroy17a,Sharda18}; see \citealt{Krumholz18a} for a recent review) 
This means that on giant molecular cloud (GMC) or molecular cloud length scales ($\sim 100$ pc down to $1$ pc), star formation is very inefficient. However, as we move towards smaller length scales ($\sim$ 0.1 pc down to AU scales) tracing gas at densities exceeding $\sim 10^7$ cm$^{-3}$, eventually there must be some density or size scale beyond which most of the mass in gas would wind up in a star $\sim 1$ dynamical time later. Therefore, there should exist a point after which $\epsilon_{\rm ff}$ does not remain small anymore and approaches unity.

Nearly every physical model of star formation predicts the existence of a threshold of this type. Knowing its value would tell us a great deal about how star formation works and what regulates it. For example, one could imagine that the dense clumps traced by HCN emission are gravitationally bound structures, whereas gas in GMCs is largely unbound, and this is why star formation is inefficient on GMC scales \citep[e.g.,][]{H10,Lada12}. If that explanation were correct, one would expect to find low $\epsilon_{\rm ff}$ on the scales of GMCs, but high $\epsilon_{\rm ff}$ in regions traced by HCN. Such an observation would be powerful evidence that a change in boundedness is what is regulating star formation. Alternately, a number of authors have proposed models in which star formation is regulated by supersonic turbulence \citep{KM05, FK12, FK13, PN11, HC11, Hopkins2012a, Hopkins2013a, F15}. A generic feature of such models is the existence of a characteristic density scale (or a range of them in some cases) at which gas becomes bound; this too represents a predicted threshold at which we would expect a change in $\epsilon_{\rm ff}$. Similar arguments can be used to derive critical densities, column densities, or length scales from magnetic-regulation models of star formation \citep{Shu1987,McKee89a,Basu04magnetic}.

\begin{figure}
    \centering
    \includegraphics[width=\columnwidth]{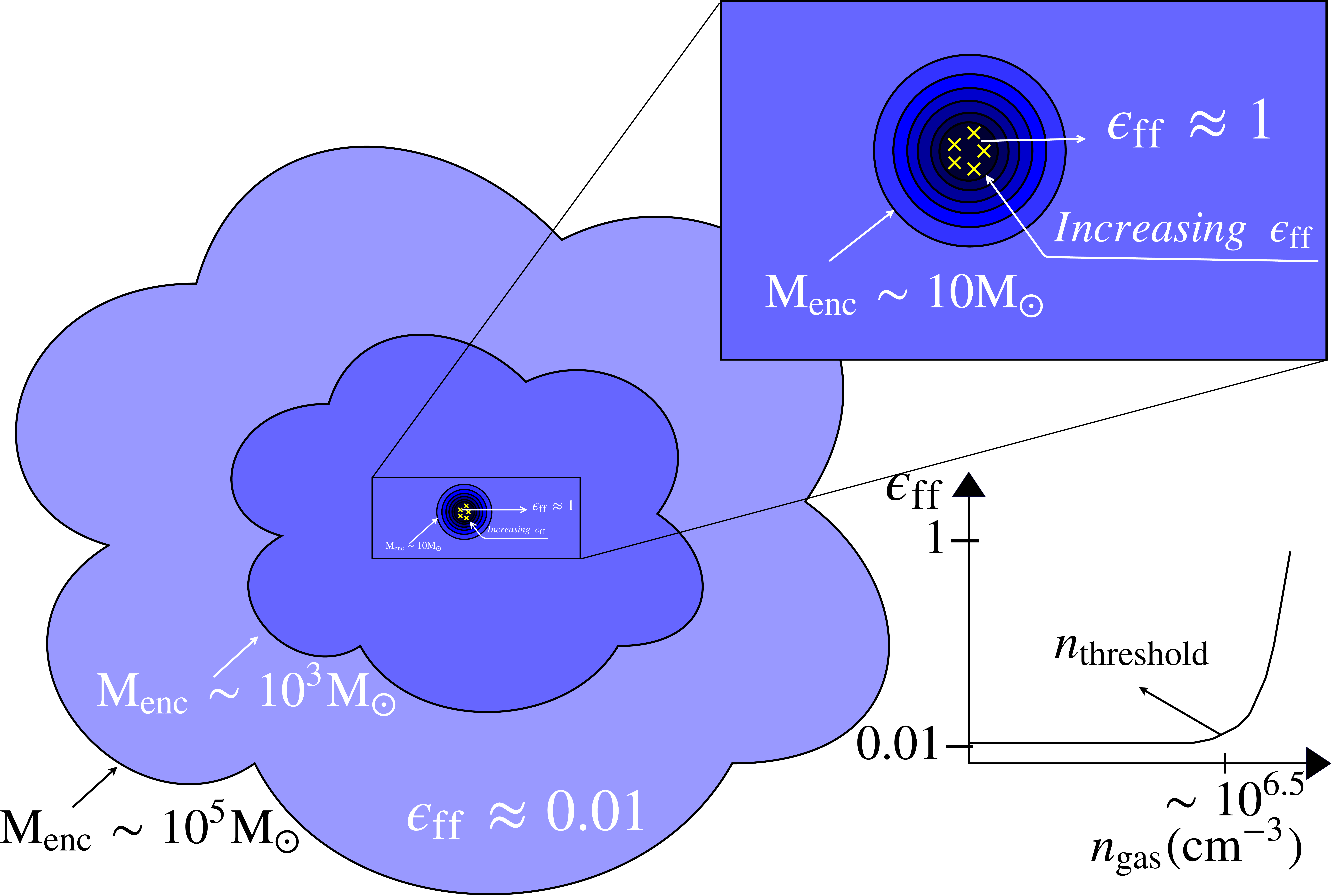}
    \caption{A pictorial description of a density threshold in a $10^5$ M$_\odot$ giant molecular cloud. The yellow crosses in the inset represent stars and the lines represent contours of increasing density. At low density $\epsilon_{\rm ff}$ is small, as depicted in the inset. Since fluid elements only increase their densities gradually on average, mass builds up and there is a great deal of mass between a pair of contours. Once one reaches a density such that $\epsilon_{\rm ff}\approx 1$, mass moves rapidly from lower to higher density, and spends little time at intermediate densities. Thus there is little mass between two density contours at any given time, and the contours are closely spaced. The transition from widely-spaced to narrowly-spaced contours marks the threshold density $n_{\rm thresh}$, which we show below for our simulations is $\sim 10^{6.5}$ cm$^{-3}$.}
    \label{fig:eff-cloud}
\end{figure}

In order to detect a threshold in observations, we must know what it looks like, and for this purpose it is helpful to envision the process of star formation in a Lagrangian sense: as a conveyor belt moving mass from low to high to stellar densities. The value of $\epsilon_{\rm ff}$ characterises the speed of this flow for any particular fluid parcel: $\epsilon_{\rm ff}\ll 1$ at a given density $\rho$ means either that fluid parcels of density $\rho$ require many free-fall times $t_{\rm ff} \approx 1/\sqrt{G\rho}$ to substantially increase in density, or that fluid parcels that are increasing in density are nearly balanced out by those decreasing in density, so the net flow of mass to higher density is small. On the other hand, values of $\epsilon_{\rm ff}\approx 1$ mean that the typical fluid parcel requires only a time $\approx t_{\rm ff}$ to go up in density. Since the mass flow from low to high density must be continuous, at least in a time-averaged sense, the value of $\epsilon_{\rm ff}$ determines how much mass is ``stuck'' at a given density along the conveyor belt: low $\epsilon_{\rm ff}$ corresponds to places where the conveyor belt moves slowly and thus large amounts of mass build up, while high $\epsilon_{\rm ff}$ corresponds to places where the conveyor belt moves rapidly and there is relatively little mass. Thus the observational signature of a threshold is as illustrated in the cartoon in \autoref{fig:eff-cloud}: if one selects a density such that $\epsilon_{\rm ff} \ll 1$, then there is a large amount of mass stuck at densities from $n$ to $2n$, and contours drawn at these two densities will enclose a great deal of mass. If one selects a higher density $n$ such that $\epsilon_{\rm ff} \approx 1$, there is little mass between $n$ and $2n$. The threshold $n_{\rm thresh}$ is the density where the contours switch from being widely-spaced to narrowly-spaced in mass.

While this manifestation is relatively straightforward to investigate in a simulation where we have access to the full 3D density field, for observed clouds where we have access to quantities only in projection, it is obviously not possible to search for a threshold as illustrated in \autoref{fig:eff-cloud}. Instead, one must use some observable proxy. \citet{Onishi98}, \citet{Johnstone04}, and \citet{Hatchell05} were among the first to search for a threshold using observations. \cite{Onishi98} studied the relations between cores and young stellar objects in Taurus, and found that C$^{18}$O cores that contained either compact dense centres traced by H$^{13}$CO or cold Infrared Astronomical Satellite (IRAS) sources have high column densities, and therefore concluded that gravitational collapse and subsequent mass accretion occurs once a core's column density exceeds $\sim 8 \times 10^{21}$ cm$^{-2}$ ($A_V \sim 7-9$ mag).
\cite{Johnstone04} studied substructures in the Ophiuchus clouds using submillimeter continuum maps and compared them with visual extinction maps. These authors note the absence of any structures for $A_V \leq 7$ mag, and argue that this indicates that substructures only form where $A_V \geq 15$ mag.

\cite{H10} and \cite{L10} lend support to the idea that there is a threshold based on data for a larger sample of local molecular clouds. They argue that the correlation between the star formation rate and gas mass becomes increasingly tight as one considers the gas mass at higher and higher column densities. \citet{Konyves15}, \citet{Andre15}, and \citet{Andre2017} find that the objects they identify as cores in \textit{Herschel} maps (which are in practice defined by contours at a certain signal to noise threshold) are found almost exclusively on background material characterized by an extinction of $A_V \sim 7-9$ mag. They argue that this extinction level is characteristic of the densities at which gravitational collapse of filamentary structures takes place. \cite{Lada12} propose that the existence of a density threshold explains why external galaxies show a near-linear correlation between star formation rate and mass of gas traced by HCN line emission, but a superlinear correlation between star formation and CO line emission. In their model, HCN traces gas above the threshold, while CO traces gas below it.

However, a number of authors have questioned these conclusions. \citet{Garcia-Burillo12a}, \citet{Usero15a}, and \citet{Bigiel16a} show that the correlation between HCN emission and star formation in extragalactic systems is not in fact linear, calling into question the idea that HCN traces gas that is especially closely linked to star formation (though see \citealt{Shimajiri17a} for an opposing view). \citet{Elmegreen18a}, following up on earlier arguments by \citet{KrumholzThompson07}, argues that the near-linearity of the HCN-star formation correlation is an observational selection effect rather than a physical threshold. For Galactic measurements, \cite{BH13} argue that the correlations observed by \citet{H10} and \citet{L10} do not require a particular density threshold, and variation in the dependence of $\Sigma_{\rm SFR}$ on $\Sigma_{\rm gas}$ is instead due to the increasing importance of gravity at higher densities. Consistent with this picture, \citet{Gutermuth11a} show that the density of young stellar objects increases as a smooth powerlaw with gas surface density, again suggesting a smooth rise in star formation rate in dense gas without any special threshold density value. \citet{CG14} find that the clouds in their simulations can still form stars at cloud-averaged densities which are lower than the ``threshold" value of $\sim 120 \ \mathrm{M_{\odot} \ pc^{-2}}$ proposed by \citet{H10} and \citet{L10} and suggest that their threshold for star formation is more likely a consequence of the star formation process, rather than a prerequiste for star formation. In other words, regions where star formation is more active and there are more YSOs present also tend to be regions where a great deal of gas has collapsed to high surface density, but the latter is not a direct cause of the former. 

In this paper, we aim to first confirm that there is a threshold density above which $\epsilon_{\rm ff}$ approaches unity in detailed simulations of star formation, and then to investigate whether we can recover this value from observations using a variety of proposed diagnostic methods. We answer these questions by analyzing simulations that include some of the most detailed descriptions of the different physical processes relevant for star formation, and mimic conditions prevalent in nearby molecular clouds. We use these simulations to create mock observations, and test whether various proposed methods for detecting thresholds in observations can in fact recover results that match what we obtain using the full 3D spatial and temporal information to which we have access for the simulations.

The rest of the paper is structured in the following manner. We begin by describing the simulations and the methodology used to create mock observations in \autoref{sec:Methodology}. In \autoref{sec:Results}, we first search for star formation thresholds using the full simulation information (in particular the gas volume density), and then assess the capability of various observational analysis methods to recover such a threshold using the projected information to which observers have access. We summarize our findings and the main conclusion of our study in \autoref{sec:Conc}.

\section{Simulations and Analysis Methods}
\label{sec:Methodology}

In our study we use the simulations described in \cite{F15} and \cite{C18} (hereafter F15 and C18, respectively). We choose these simulations since they include detailed treatments of gravity, turbulence, magnetic fields, mechanical jet/outflow and radiation feedback and obtain star formation rates (F15) and IMFs (C18) that are among the closest matches to observations to date. We briefly highlight their main features strictly relevant to this work and refer the reader to F15 and C18 for more details. We also summarize some key simulation parameters in \autoref{tab:Sim-params}.

\subsection{F15 Simulations}
\label{sec:F15sim}

F15 use the AMR code \citep{Berger1989} {\tt{FLASH}} \citep{Fryxell2000, Dubey2008} to solve the compressible MHD equations. While the simulations described in F15 include physical processes in steps of increasing complexity, we only use the one with the most complete set of physical processes, which includes self-gravity, turbulence, magnetic fields, and jet/outflow feedback. The simulation that we use here is not directly from F15, but is constructed with the same initial and boundary conditions as all the simulations in F15, is identical to the most complex simulation in F15 labelled GvsTMJ and additionally includes radiation feedback based on the implementation by \cite{FKH17} who use the model described in \citet{Offner2009} for protostellar evolution. This simulation is labelled GTBJR in \cite{Onus18} and we adopt the same name for the purposes of this work. The turbulence driving in the simulation excites a natural mixture of solenoidal and compressible modes, corresponding to a turbulence driving parameter $b = 0.4$ \citep{Federrath2010}. We refer the reader to F15 for more details on the implementation of sink particle formation, turbulence, magnetic fields, jets/outflows \citep{FederrathEtAl2014} and to \citet{FKH17} for the implementation of radiation feedback. Fragmentation, star formation and accretion are modelled with the sink particle technique by \citet{FederrathEtAl2010}.

The simulation box has a total cloud mass $M = 388 \ \mathrm{M}_{\odot}$ and has a box length of $L = 2$ pc, with a mean density of $\rho_0 = 3.28 \times 10^{-21}$ g cm$^{-3}$, corresponding to a global free-fall time of $t_{\rm ff} = 1.16$ Myr. The velocity dispersion is $\sigma_v = 1$ km s$^{-1}$, and the sound speed is $c_s = 0.2$ km s$^{-1}$ at the initial temperature $T=10$ K, which results in an rms Mach number of $\mathcal{M} = 5$. The simulation starts with an initial uniform magnetic field of $B=10$ ${\rm \mu}$G. These simulations can be characterized by the magnetic field strength parameter $\mu_{\Phi}$, constant throughout for the whole cloud (albeit varying locally within the cloud) since mass and magnetic flux are conserved for the entire simulation box, defined as 

\begin{equation}
\label{eq:mu-phi}
    \mu_{\Phi} = \frac{M}{M_{\Phi}} = 2\pi \sqrt{G} \left( \frac{M}{\Phi} \right),
\end{equation}
where $M_{\Phi} = \Phi/2\pi \sqrt{G}$ is the magnetic critical mass (\citealt{Mouschovias1976}; the mass below which a cloud cannot collapse and above which collapse cannot be prevented by magnetic fields alone), and $\Phi$ is the total magnetic flux threading the cloud. The value of $\mu_{\Phi}$ for this simulation is 3.8. The resulting virial ratio is $\alpha_{\rm vir} = 1.0$ and the plasma beta is $\beta = 0.33$ (corresponding to an Alfv{\'e}n Mach number of $\mathcal{M}_{\rm A} = 2.0$). These physical conditions are chosen to mimic those found in nearby, low-mass, star-forming regions such as Perseus or Taurus.

Sink particles in the simulation are formed dynamically when a local region undergoes gravitational collapse. Once the gas density in a cell exceeds a density of $\rho_{\rm sink} = \pi c_s^2/G \lambda_{\rm J}^2$, a control volume of radius $r_{\rm sink} = \lambda_{\rm J}/2$ is formed around it and it is checked whether all the gas in that volume is Jeans-unstable, is gravitationally bound and is collapsing towards the central cell. If all these additional checks are passed, a sink particle is formed in the central cell. Performing these additional checks suppresses spurious sink formation in transient shocks \citep{FederrathEtAl2010}.

\subsection{C18 Simulations}
\label{sec:C18sim}

C18 use the {\tt ORION2} AMR code \citep{Lietal2012} to solve the equations of ideal MHD along with treatments of coupled self-gravity \citep{Truelove1998,Klein1999}, and radiation transfer \citep{Krumholzetal2007}. The simulations also include feedback due to protostellar outflows following the procedure in \citet{Cunningham2011}; stars form following the sink particle algorithm of \citet{Krumholz04}, and protostellar evolution uses the model described in \citet{Offner2009}. The gas is initially evolved under the action of a turbulent driving force for two crossing times, $t_{\rm cross} = L/v_{\rm rms} = 0.51$ Myr, after which self gravity is switched on. The evolution of the system thereafter is categorized into two cases, one where turbulence is allowed to decay and the other where a constant rate of energy is injected to balance the rate of turbulent decay \citep{MacLow1999}. Although C18 carry out simulations with a wide range of $\mu_{\Phi}$ values, we focus on the simulations with $\mu_{\Phi}$ = 1.56 because this value is comparable to observed mass to flux ratios in nearby molecular clouds, and because these simulations run long enough to produce enough protostars to yield meaningful statistics. These simulations are also inefficient in forming stars for the driven turbulence case. We label the two cases C18Decay and C18Drive (based on whether turbulence is allowed to decay or being driven) for the purposes of this work; these correspond to the runs in Rows 1 and 3 in Table 1 of C18. For more details of the physics included and the limitations of the simulations, we refer the reader to C18.

The initial temperature and sound speed are the same as for the F15 simulations. The turbulent driving force is purely sinusoidal (b$\sim$0.33) and is scaled to maintain an rms Mach number of $\mathcal{M}=6.6$, leading to a virial parameter $\alpha_{\rm vir} = 1.05$. The simulation box has a total cloud mass of $M$ = 185 $\mathrm{M}_{\odot}$ and has a box length of $L = 0.65$ pc, with a mean density of $\rho_0 = 4.46 \times 10^{-20}$ g cm$^{-3}$ corresponding to a global free-fall time of $t_{\rm ff} = 0.315$ Myr. The value $\mu_\Phi = 1.56$ corresponds to a magnetic field of $B=81.3$ ${\rm \mu}$G and an Alfv{\'e}n  Mach number of $\mathcal{M}_{A}$ = 1.4.

Sink particles in the simulation are formed only on the finest AMR level when the gas becomes dense enough to exceed a local Jeans number of $J = (G \rho \Delta x^2/\pi c_s^2)^{1/2} >1/4$ or equivalently $\rho > \rho_{\rm sink} = \pi c_s^2/16 G \Delta x^2$, where $\Delta x$ is the cell width at the finest AMR level.

\begin{table*}
	\centering
	\caption{Column 1: Simulation name. Columns 2-4: type of turbulence driving,  3D velocity dispersion, rms Mach number. Columns 5-6: ratio of thermal to magnetic pressure ($\beta$) and Alfv{\'e}n Mach number. Column 7: Dimensionless mass to flux ratio (\autoref{eq:mu-phi}). Columns 8-10: length of the simulation box, mean density, the sink particle threshold density. Columns 11-12: Time at which the simulation snapshot is taken and total number of stars present in the simulation snapshot used. Column 13: SFR per mean global freefall time ($\epsilon_{\rm ff}$).}
	\begin{tabular}{lcccccccccccr} 
		\hline
		Name & Turb.~($b$) &  $\sigma_v$ & $\mathcal{M}$ & $\beta$ & $\mathcal{M_{\rm A}}$ & $\mu_{\Phi}$ & $L$ & $\rho_0$ & $\rho_{\rm sink}$ & Time & $N_{\rm stars}$ & $\epsilon_{\rm ff}$\\
		& & & [km/s] & & & & [pc] & [g cm$^{-3}$] & [g cm$^{-3}$] & [Myr] & &  \\
		(1) & (2) & (3) & (4) & (5) & (6) & (7) & (8) & (9) & (10) & (11) & (12) & (13) \\
		\hline
		GTBJR  & Mix ($b$=0.4)  & 1.0 & 5.0 & 0.33 & 2.0 & 3.8 & 2 & 3.28 $\times$ $10^{-21}$ & 8.3 $\times$ 10$^{-17}$  & 4.254 & 12 & 0.031 \\
		C18Decay  & None  & 1.254  & 6.6 & 0.046 & 1.0 & 1.56 & 0.65 & 4.46 $\times$ $10^{-20}$ & 1.16 $\times$ 10$^{-15}$  & 1.91 & 55 & 0.12\\
	C18Drive & Sol.~($b$=0.33) & 1.254 & 6.6 & 0.046 & 1.0 & 1.56 & 0.65 & 4.46 $\times$ $10^{-20}$ & 1.16 $\times$ 10$^{-15}$  & 1.842 & 63 & 0.048\\
		\hline
	\end{tabular}
	\label{tab:Sim-params}
\end{table*}

\subsection{Creating Mock Observations}

To create mock observations of column density we choose snapshots from GTBJR, C18Decay and C18Drive at times corresponding to 4.254, 1.91 and 1.842 Myr respectively and make projection maps along each of the three cardinal axes. To compare these maps to observations of nearby molecular clouds, we smooth them with a Gaussian kernel with a FWHM corresponding to typical resolutions in observations using dust-based tracers; we focus on dust rather than molecular lines because dust measurements generally provide the most accurate estimates of total column density on small scales \citep{Goodman2009}. There are a wide range of distances and resolutions found in observational studies available in the literature. The physical resolution depends on target distance, observation wavelength, instrument, and technique. For this reason we consider two representative cases: (i) a resolution of 3.0 arcmin (typical of NIR extinction measurements; \citealt{NICESTRes2016}) for a cloud at a distance of 140 pc (typical of Taurus; \citealt{Taurus}), corresponding to an absolute resolution of 0.070 pc and (ii) a resolution of 36.9 arcsec (typical for \textit{Herschel} observations at 500 $\mu$m; \citealt{HerschelRes2010}) at a distance of 260 pc (typical of Aquila; \citealt{Aquiladist2003,Konyves15}), corresponding to an absolute resolution of $\approx$ 0.046 pc. We shall denote these two resolutions as Res1 and Res2 respectively.

\begin{figure}
    \centering
    \includegraphics[width=\columnwidth]{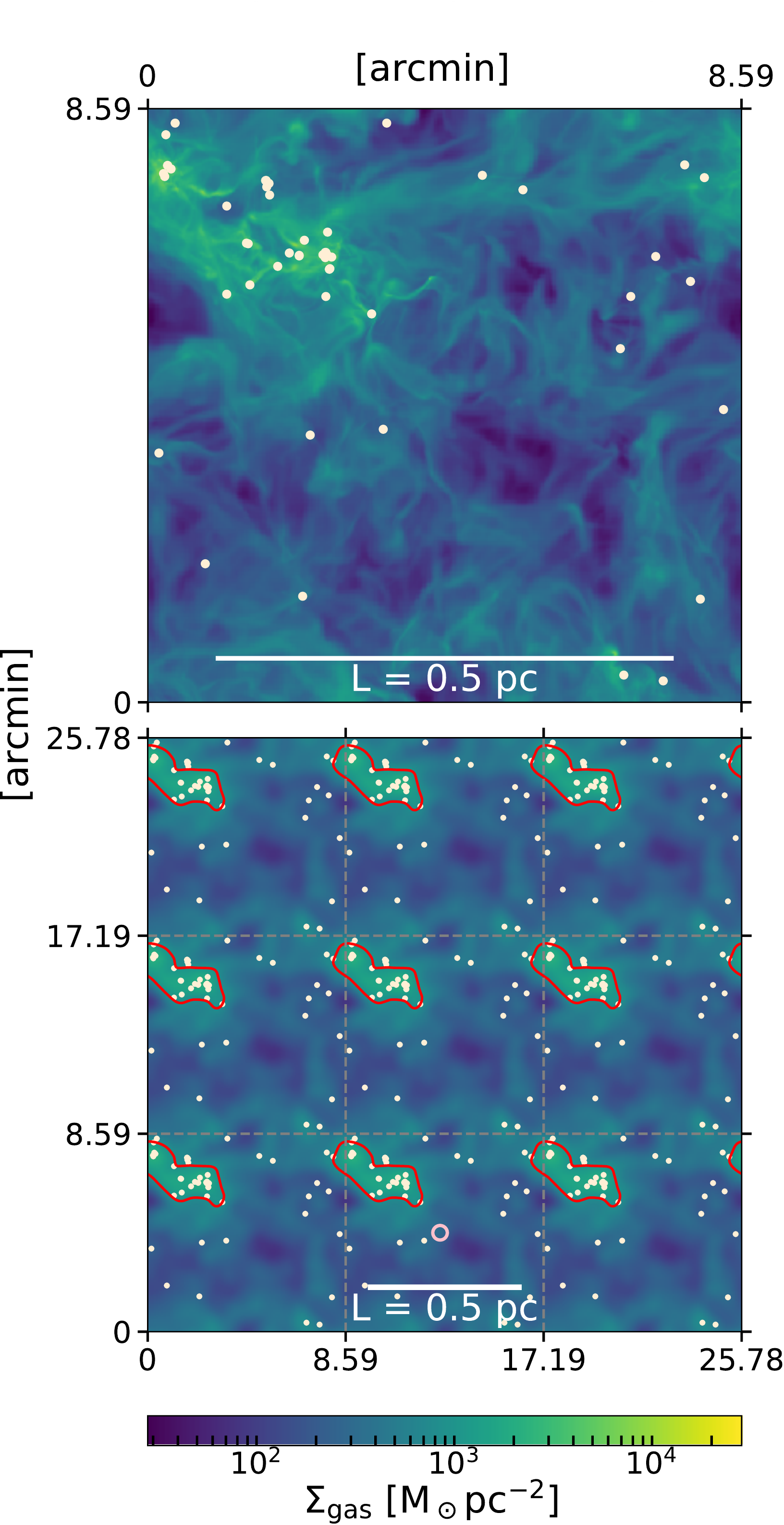}
    \caption{
    \label{fig:method}
    \textit{Top}: Column density map for the simulation C18Drive at a time of 1.842 Myr. \textit{Bottom}: Mock observations of the top panel. Since the simulation uses periodic boundary conditions, we have tiled the box to create an image of 3072 $\times$ 3072 pixels from the raw simulation resolution of 1024 $\times$ 1024 pixels (central square, top figure). In both panels the angular scale on the sides is for a cloud at a distance of 260 pc. The red contours correspond to $\Sigma_{\rm gas} = 900$ M$_\odot$ pc$^{-2}$. White circles indicate the positions of stars. The pink circle in the lower center represents the smoothing kernel used to create this image, corresponding to a FWHM of 36.9 arcsec (Res2).}
\end{figure}

We draw contours on the resulting smoothed image at specific surface density values (which we also refer to as ``contour levels''), $\Sigma_{\rm gas}$, evenly spaced throughout the entire range available for a given projection map along each axis. Since the contour shapes can vary and some contours can stretch over the entire length of the simulation box, we use only the ones that form closed curves. In cases where there are multiple closed contours for a given contour level, we combine the areas of the different unique contours and treat them as a single entity. We then project the positions of all the stars (sink particles) formed in the simulations onto these maps and count the number of Young Stellar Objects (YSOs), i.e., stars having an age less than 0.5 Myr (Class 0/I), enclosed within each contour (denoted by $N_{\rm YSO}$). We compute the following quantities for each contour level on both the smoothed and un-smoothed maps: (i) the enclosed area ($A$), (ii) the enclosed gas mass ($M_{\rm gas}$), (iii) the total mass of the YSOs ($M_*$), and (iv) the free-fall time 
\begin{equation}
\label{eq:tff}
    t_{\rm ff,2D} = \sqrt{\frac{3 \pi}{32 G \rho_{\rm 2D}}},
\end{equation}
where $\rho_{\rm 2D} = 3 \sqrt{\pi} M_{\rm gas}/4 A^{1.5}$ is the mean density that would be estimated by an observer under the assumption that the line-of-sight size of the region is comparable to the size projected on the plane of the sky \citep[e.g.,][]{KDM12}. We show an example of the method in \autoref{fig:method}. 

\section{Results}
\label{sec:Results}

Having generated our simulated observations, we now investigate how well one can diagnose star formation thresholds from them using various techniques. We begin in \autoref{sec:eff-rho-gas} by investigating the physical threshold in the simulation using the full 3D and time-dependent information to which we have access. In the remainder of this section we investigate various methods for trying to recover this information from 2D projected data and mock observations.

\subsection{The true threshold: variation of $\mathbf{\epsilon_{\rm ff}}$ with $\mathbf{\rho_{\rm gas}}$}
\label{sec:eff-rho-gas}

As discussed in \autoref{sec:Intro}, the dimensionless quantity $\epsilon_{\rm ff}$ is the fraction of an object's gas mass that is transformed into stars in one free-fall time at the object's mean density.  Since denser regions invariably form stars more quickly than more diffuse ones (indeed, it would be very surprising if they did not), any claim of any type of volume or column density ``threshold'' for star formation below which star formation is suppressed, must reveal itself as a significant change in $\epsilon_{\rm ff}$ across the threshold. 

\begin{figure}
    \centering
    \includegraphics[width = \columnwidth]{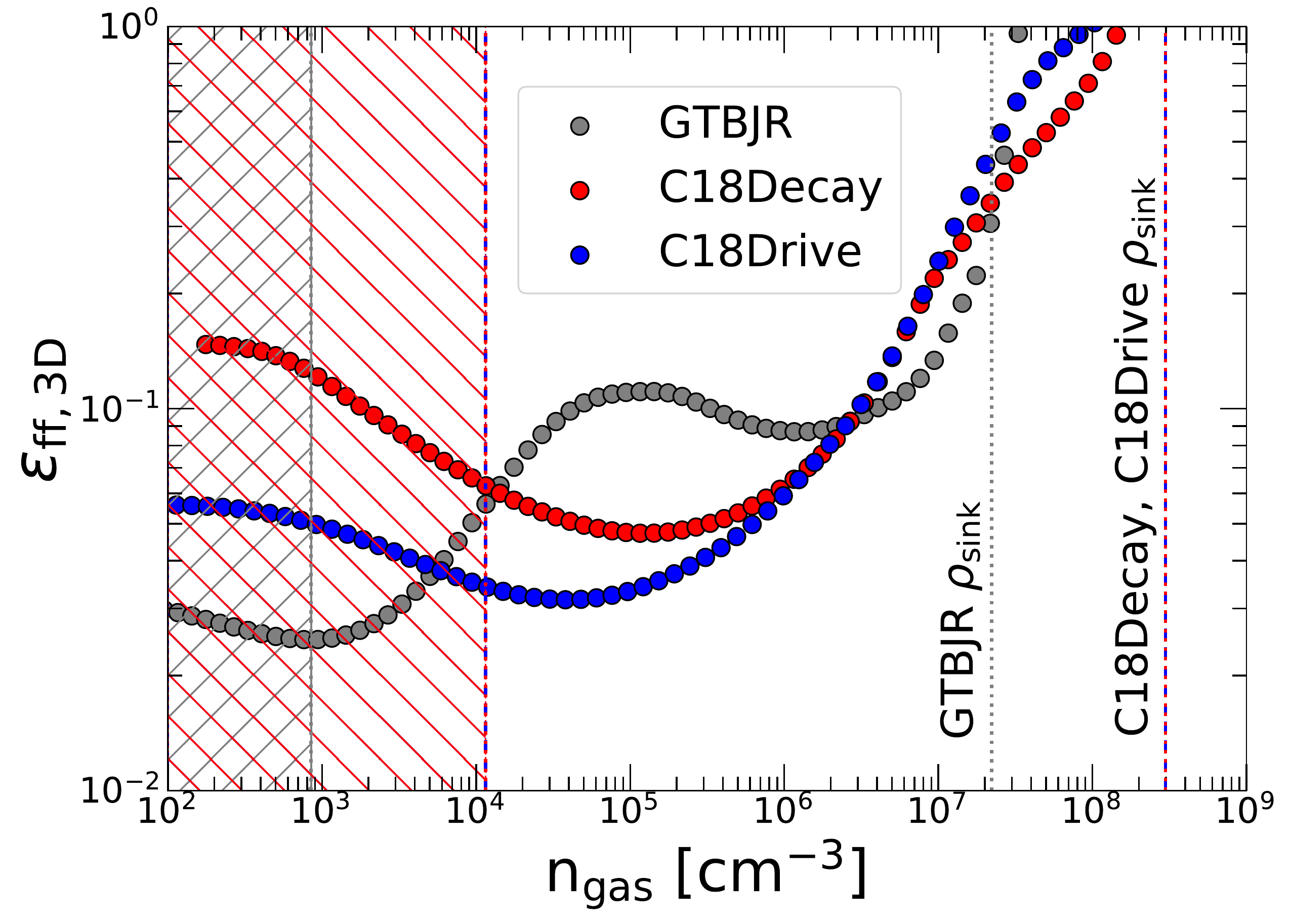}
    \caption{
Star formation rate per free-fall time $\epsilon_{\rm ff,3D}$ (\autoref{eq:epsff3d}) as a function of gas density. The vertical lines are the sink particle creation thresholds in the F15 and the C18 simulations, as indicated. The hatched regions indicate the data points below the mean density where $\epsilon_{\rm ff}$ rises. This rise is an artifact of the periodic box used in the simulations (see text).}
    \label{fig:eff-rho-gas}
\end{figure}

At a minimum the simulations must have a threshold because they include sink particles -- by construction gas that meets the conditions for sink particle creation $\epsilon_{\rm ff} \sim 1$. A necessary but not sufficient condition for sink particles to form is that the density exceed a threshold value $\rho_{\rm sink}$ (or number density $n_{\rm sink}$), which is $2.13 \times 10^7$ cm$^{-3}$ for GTBJR and $2.9 \times 10^8$ cm$^{-3}$ for C18Decay and C18Drive. However, a transition from low efficiency to free-fall collapse could also occur at densities considerably below these values. To determine if this is the case, in \autoref{fig:eff-rho-gas} we plot $\epsilon_{\rm ff}$ as a function of $\rho_{\rm gas}$ using 
\begin{equation}
\label{eq:epsff3d}
    \epsilon_{\rm ff, 3D} = \frac{t_\mathrm{{ff}} (\rho') }{M_{\mathrm{gas,>}}}  \times  \mathrm{SFR}  ,
\end{equation}
where SFR is the true, time-averaged star formation rate in the simulations, $M_{\mathrm{gas,>}}$ is the mass of gas above a certain density value chosen as the threshold ($\rho_{\rm gas, th}$), and $t_\mathrm{{ff}(\rho')}$ is the free-fall time evaluated at $\rho'$, which is the mean density for all the gas with $\rho >\rho_{\rm gas, th}$. This method is equivalent to 3D contouring (the 3D version of the method used in \autoref{sec:Methodology}) with the assumption that all the stars lie in the densest regions.

\autoref{fig:eff-rho-gas} shows a few interesting features. $\epsilon_{\rm ff, 3D}$ is roughly constant to within a factor of $\sim 3$ from $n\approx 10^4 - 10^{6.5}$ cm$^{-3}$ for all the three simulations. (The GTBJR simulation also shows a slight bump in $\epsilon_{\rm ff}$ at $n\sim 10^5$ cm$^{-3}$, but this appears to be a transient feature; similar features appear and disappear at a range of intermediate $n$ values in other F15 simulations (not shown), but do not appear at a consistent density or time, and no similar features appear in the C18 simulations.) Note that, while $\epsilon_{\rm ff}$ is constant from $n\approx 10^4 - 10^{6.5}$ cm$^{-3}$, it begins to rise for even lower densities, particularly in the C18 simulations. We disregard this effect because it is an artifact of the periodic box used in the simulations. For density thresholds approaching the mean simulation box density ($\approx 10^4$ cm$^{-3}$ for C18, $\approx 10^3$ cm$^{-3}$ for GTBJR), further decreases in density do not add any more mass or star formation inside the contour. It is this effect that drives the rise in $\epsilon_{\rm ff}$ at the lowest densities.

Despite these artifacts at low density, \autoref{fig:eff-rho-gas} provides clear evidence of a physical threshold: $\epsilon_{\rm ff, 3D}$ is roughly constant and $\lesssim 0.1$ over the range from $n\approx 10^3 - 10^{6.5}$ cm$^{-3}$, then rises sharply at $n \approx 10^{6.5}$ cm$^{-3}$. This is clearly well below the threshold density at which sink particles form,\footnote{Note that, for GTBJR, due to the large number of additional checks (see e.g. \citet{FederrathEtAl2010}) imposed before sink particles form, the mean density at which sink particles form in practice is considerably higher than $\rho_{\rm sink}$.} indicating that this rise is not due to the sink particle algorithm. Instead, \autoref{fig:eff-rho-gas} provides clear evidence that star formation in the simulations transitions from inefficient, $\epsilon_{\rm ff} < 0.1$, to efficient, $\epsilon_{\rm ff} \sim 1$, at $n \approx 10^{6.5}$ cm$^{-3}$. Thus, we confirm that there exists a true volume density threshold for star formation in the simulations. We therefore next investigate how well various observational methods can recover it.

\subsection{Thresholds in projection: $\mathrm{\epsilon_{\rm ff}}$ vs.~$\mathbf{\Sigma_{\rm gas}}$ }
\label{sec:eff-sigma-gas}

In this section, we consider a first possible method to recover the volume density threshold using essentially the same method proposed by \citet{KDM12}, \citet{Federrath2013}, \citet{SFK15}, and \citet{Sharda18}, which simply represents a projected version of the volumetric analysis used above.

\begin{figure}
    \centering
    \includegraphics[width=\columnwidth]{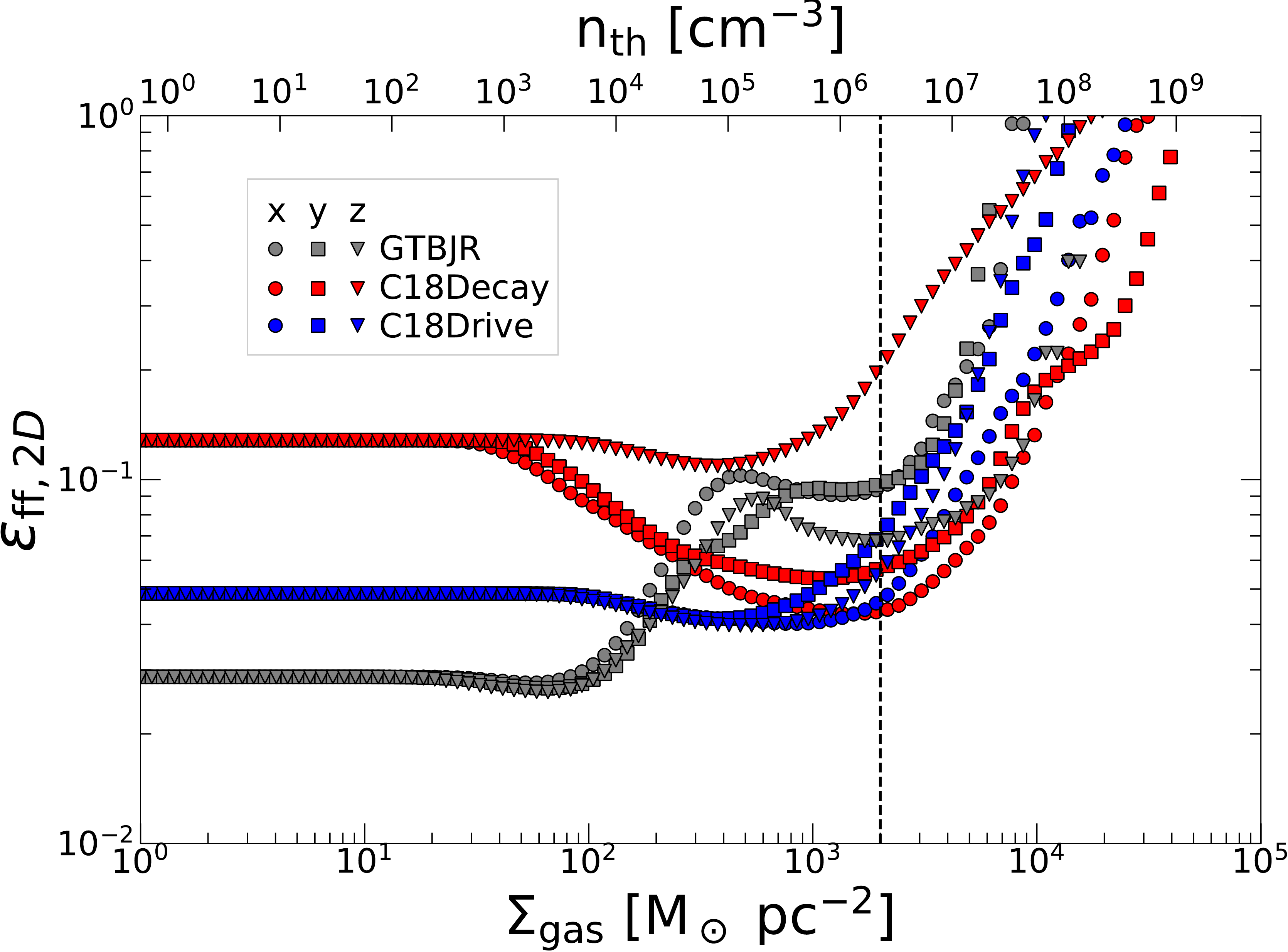}
    \caption{
    Projected estimate of star foramtion rate per free-fall time $\epsilon_{\rm ff,2D}$ (see text) versus contour level $\Sigma_{\rm gas}$ for the un-smoothed projected density maps. Different shapes correspond to projections along the three different cardinal axes. The top axis indicates the corresponding $n_{\rm th}$ estimated from the fiducial model (\autoref{eq:rhoth}) and the vertical dashed line serves to facilitate comparison between the surface and volume density threshold values.}
    \label{fig:eff2d}
\end{figure}

We first estimate $\epsilon_{\rm ff}$ (denoted $\epsilon_{\rm ff,2D}$ here, to distinguish the projected, 2D from the true, 3D value) for each contour level in the un-smoothed column density maps for each simulation (see e.g. top panel in \autoref{fig:method}) as

\begin{equation}
\epsilon_{\rm ff,2D} = \frac{t_{\rm ff,2D}}{M_{\rm gas}} \times  \mbox{SFR},
\end{equation}
where $t_{\rm ff,2D}$ is the free-fall time estimated from projected quantities, per \autoref{eq:tff}. \autoref{fig:eff2d} shows the variation of $\epsilon_{\rm ff}$ as a function of contour level $\Sigma_{\rm gas}$ for this un-smoothed image. The curves are qualitatively similar to those shown in \autoref{fig:eff-rho-gas} and show the ``rising $\epsilon_{\rm ff}$'' feature at the highest densities. Therefore, evidence for a density threshold can be recovered from such a plot. To estimate the threshold volume density $\rho_{\rm th}$ associated with the column density $\Sigma_{\rm th}$ where $\epsilon_{\rm ff,2D}$ begins to rise, we can hypothesise that the characteristic line of sight depth should be of order the Jeans length, i.e.,
\begin{equation}
\label{eq:rhoth}
    \Sigma_{\mathrm{th}} \sim \rho_{\mathrm{th}} \ \lambda_{\rm J} (\rho_{\mathrm{th}}) \implies \rho_{\mathrm{th}} \sim \frac{\Sigma_{\mathrm{th}}^2 G}{\pi c_s^2},
\end{equation}
where $\lambda_{\rm J} (\rho_{\mathrm{th}})$ is the Jeans length evaluated at $\rho_{\mathrm{th}}$. The top axis in \autoref{fig:eff2d} shows the values of $n_{\rm th}$ evaluated using \autoref{eq:rhoth}. We find that $\rho_{\mathrm{th}}$ estimated from \autoref{eq:rhoth} and \autoref{fig:eff2d} is reasonably close to the actual values found in \autoref{fig:eff-rho-gas}, i.e., $\epsilon_{\rm ff,2D}$ begins to rise at a projected column density $\approx$ $2 \times 10^3 \ \mathrm{M}_\odot$ pc$^{-2}$, and this corresponds to an estimated volume density $n_{\rm th}\sim 10^{6.5}$ cm$^{-3}$, not far off the 3D value one would infer directly from \autoref{fig:eff-rho-gas}. 

\begin{figure}
    \centering
    \includegraphics[width=\columnwidth]{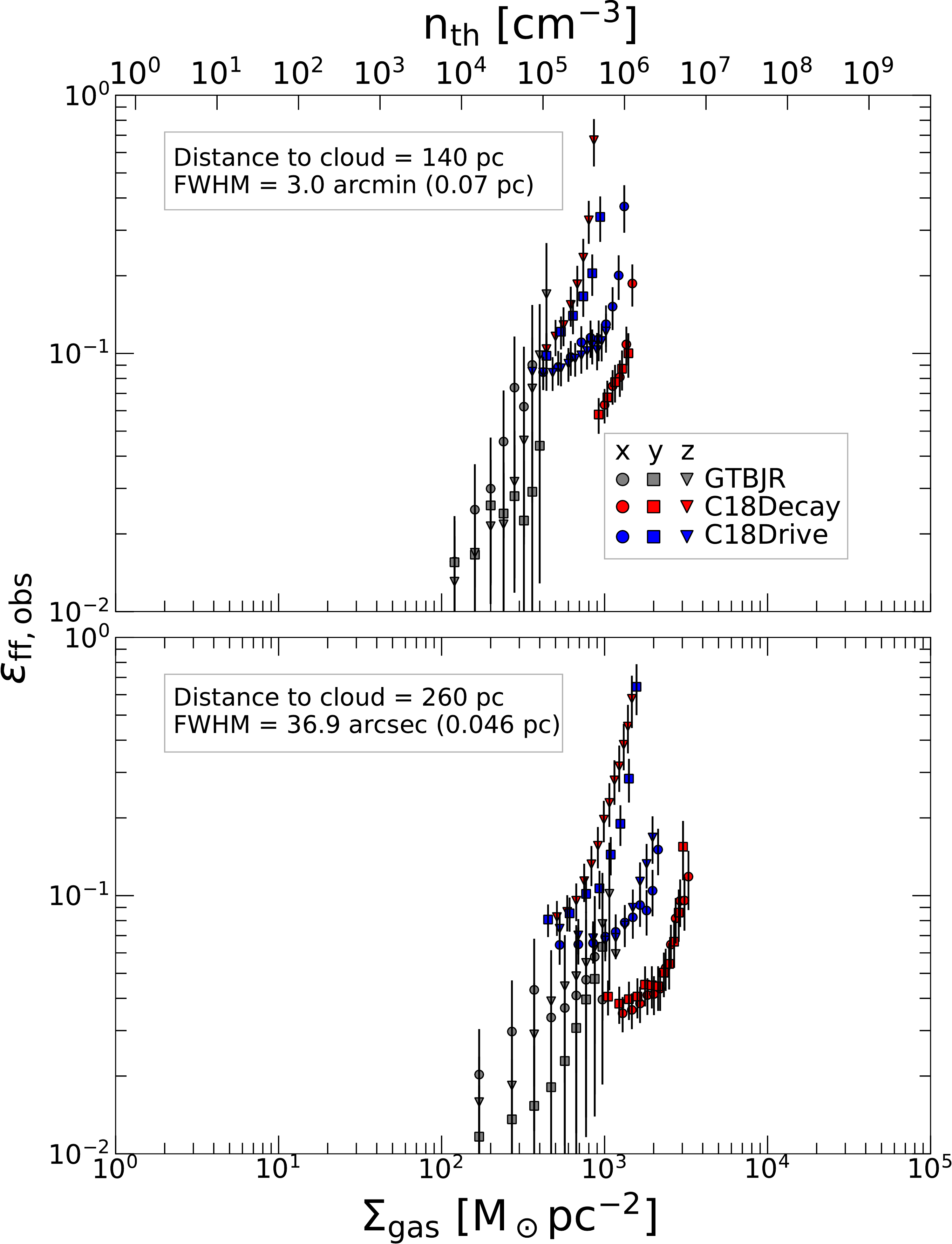}
    \caption{
    Observationally-inferred star formation rate per free-fall time $\epsilon_{\rm ff, obs}$ (\autoref{eq:eff1}) versus contour level $\Sigma_{\rm gas}$ for the mock observations (\autoref{sec:Methodology}) at two different resolutions: Res1 (\textit{top}) and Res2 (\textit{bottom}). The error bars correspond to the $\sqrt{N_{\rm YSO}}$ poissonian errors. As in \autoref{fig:eff2d}, the different shapes correspond to projections along the three cardinal axes. The top axis indicates the corresponding $n_{\rm th}$ estimated from the fiducial model (\autoref{eq:rhoth}).}
    \label{fig:eff-sigma-gas}
\end{figure}

Next we investigate whether this ``rising $\epsilon_{\rm ff}$'' feature can be recovered from the mock observations (see \autoref{sec:Methodology}) we use in this work. As is common in observational work, we use the areas $A$, gas mass $M_{\rm gas}$, YSO count $N_{\rm YSO}$, free-fall time $t_{\rm ff}$ of each contour to determine $\epsilon_{\rm ff}$ as
\begin{equation}
\label{eq:eff1}
\epsilon_{\rm ff,obs} = \left(\frac{0.5\, \mathrm{M}_\odot N_{\rm YSO}}{M_{\rm gas}}\right) \left(\frac{t_{\rm ff,2D}}{t_{\rm YSO}}\right)
\end{equation}
where $t_{\rm ff,2D}$ is again computed from \autoref{eq:tff}, but this time for the smoothed maps, and $t_{\rm YSO} = 0.5$ Myr, our nominal estimate of the class 0/I lifetime. \autoref{fig:eff-sigma-gas} shows the $\epsilon_{\rm ff}$ versus $\Sigma_{\rm gas}$ for the mock observations at the two different resolutions, Res1 and Res2. We do not probe extremely low surface density values in the mock observations since such contour levels do not possess any closed curves (contours) in the projection map. Therefore, there may be variation in the estimated value of $\epsilon_{\rm ff}$ for surface density values below the first closed contour. We see that the rising $\epsilon_{\rm ff}$ feature can still be distinguished, and thus even in the mock observations we can detect clear evidence for a threshold using this method; however, the value of the inferred threshold column density, and thus the volume density that one would infer using \autoref{eq:rhoth}, depends strongly on resolution for the two resolutions we have explored. For the lower resolution case Res1, $\epsilon_{\rm ff,obs}$ appears to begin rising at columns of several hundred M$_\odot$ pc$^{-2}$, whereas in the un-smoothed map the rise is at several times $10^3 \ \mathrm{M}_\odot$ pc$^{-2}$. The corresponding volume density threshold one would infer from \autoref{eq:rhoth} is thus close to two orders of magnitude too small. The error is much smaller for the Res2 case. We conclude that plots of $\epsilon_{\rm ff,obs}$ versus $\Sigma_{\rm gas}$ are a reliable method for detecting the existence of thresholds, but that inferred threshold values can be strongly affected by beam-smearing, which washes out the small, high-density structures one would need to measure in order to infer the threshold column density reliably.

\subsection{Thresholds from the Kennicutt-Schmidt relation: \textbf{$\Sigma_{\rm SFR}$ vs. $\Sigma_{\rm gas}$}}
\label{sec:Heidermann}

\begin{figure}
    \centering
    \includegraphics[width=\columnwidth]{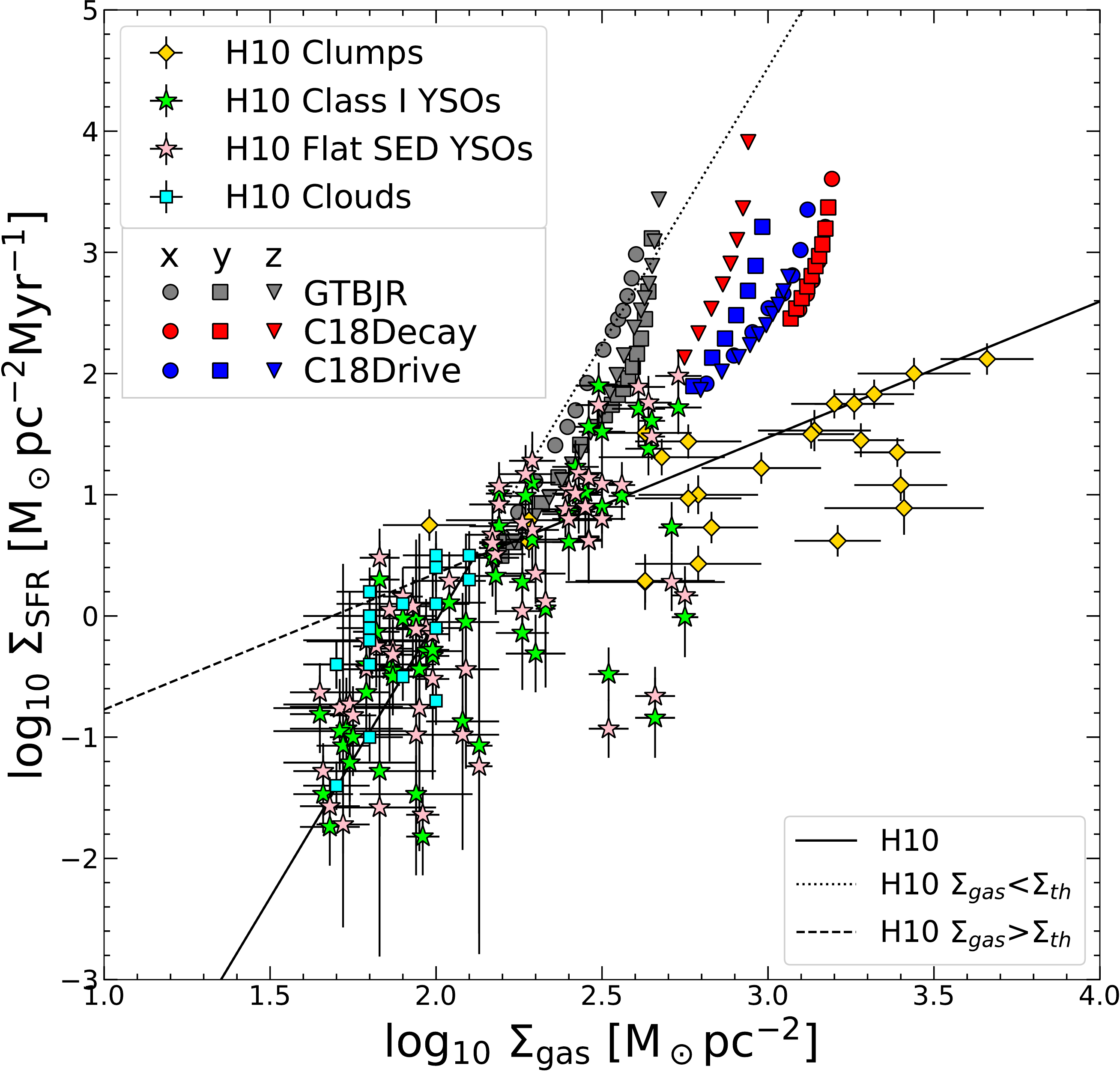}
    \caption{
    Surface density of gas $\Sigma_{\rm gas}$ interior to a given contour versus surface density of star formation $\Sigma_{\rm SFR}$ inside that contour, computed for our mock observations at resolution Res1, plotted alongside data from \citet{H10}. Different symbols and colours indicate different simulations and projection axes, as indicated in the legend. The solid black lines are H10's linear fits to the data at $\Sigma_{\rm gas}<130 \ \mathrm{M}_\odot$ pc$^{-2}$ and $>130 \ \mathrm{M}_\odot$ pc$^{-2}$, respectively; the dashed and dotted lines are the extensions of these linear fits.}
    \label{fig:KS}
\end{figure}

We next examine the method for detecting column density thresholds proposed by \citet[hereafter H10]{H10}. H10 propose searching for a threshold for star formation by plotting a Kennicutt-Schmidt (KS) relation, i.e., by plotting $\Sigma_{\rm SFR}$ versus $\Sigma_{\rm gas}$, for individual star-forming clumps. They argue that this relationship turns over from superlinear to near-linear at a surface density $\Sigma_{\rm th} \approx 129 \  \mathrm{\mathrm{M}_{\odot}}$ pc$^{-2}$, and that this provides evidence for a threshold star formation surface density. To test this method on our simulations, we compute the observationally-estimated star formation rate per unit area for the different column density contour levels as
\begin{equation}
\label{eq:SigmaSFR}
    \Sigma_{\rm SFR} =  \frac{0.5 \ \mathrm{\mathrm{M}_{\odot} N_{\rm YSO}}}{t_{\rm YSO} A},
\end{equation}
where our method of estimating the star formation rate per unit area is identical to that used by H10. We similarly compute the mean surface density of the gas inside each contour, $\Sigma_{\rm gas} = M_{\rm gas}/A$.

In \autoref{fig:KS}, we plot the relationship between $\Sigma_{\rm gas}$ and $\Sigma_{\rm SFR}$ from our simulations for the lower resolution Res1 alongside the data from H10. Changing the resolution to Res2 does not result in a qualitative change in the features of the plot. While all three simulations, GTBJR, C18Decay and C18Drive extend significantly above the observations at higher column densities, GTBJR substantially overlaps the observations at lower column densities. The discrepancy at higher columns is not all that surprising, since H10's data represent measurements only for whole ``clouds'' (defined as the outermost detectable contours given their sensitivity), not sub-regions within them as do our high column density lines. 

The main point to take from \autoref{fig:KS} is that it provides no evidence for a volume or column density threshold for star formation in our simulations, despite the fact that there is one, as is apparent from examining Figures \ref{fig:eff-rho-gas} and \ref{fig:eff2d}. To the extent that the observed data shown in \autoref{fig:KS} provide evidence for any change in star formation behaviour above a column density threshold, as opposed to being purely an observational artefact \citep{KDM12}, this threshold does not appear to be related to a change in star formation efficiency.

\subsection{Thresholds from the star to gas ratio: $N_{\rm YSO}/M_{\rm gas}$ vs. $\Sigma_{\rm gas}$}
\label{sec:Lada}

\begin{figure}
    \centering
    \includegraphics[width=\columnwidth]{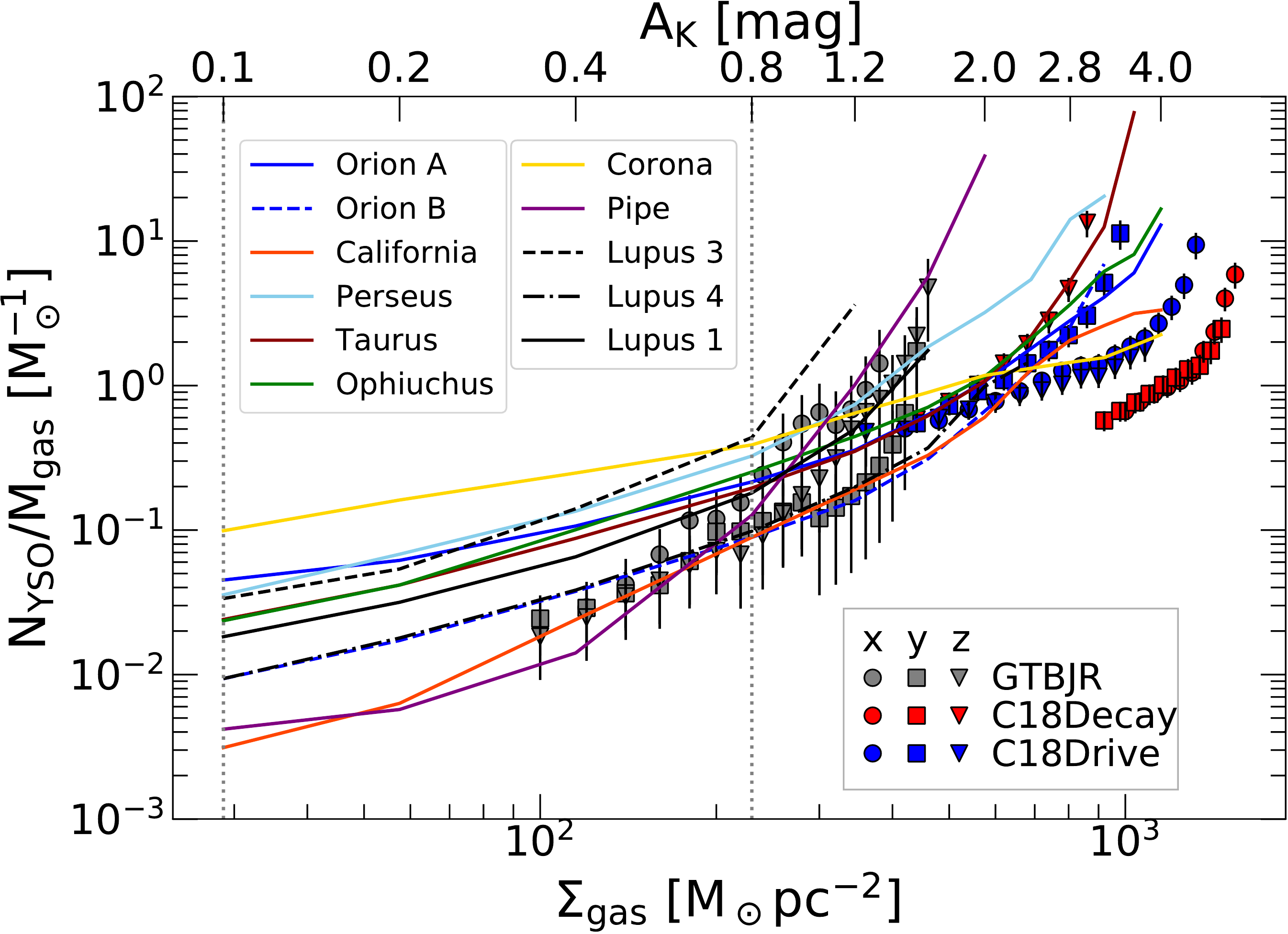}
    \caption{The star to gas ratio $N_{\rm YSO}/M_{\rm gas}$ as a function of surface density of gas $\Sigma_{\rm gas}$ from our mock observations at resolution Res1, over-plotted with data from \citet[their Figure 3]{L10}. For the mock observations, $\Sigma_{\rm gas}$ is the contour level and $N_{\rm YSO}$ and $M_{\rm gas}$ are obtained from the methodology described in \autoref{sec:Methodology}. The error bars correspond to the $\sqrt{N_{\rm YSO}}$ poissonian errors. The dotted lines at $A_\mathrm{K} = 0.1$ and 0.8 mag are the thresholds proposed by \citet{L10}. We map $A_{\rm K} = 0.8$ mag to $\Sigma_{\rm gas} = 230 \ \mathrm{M}_{\odot}$ pc$^{-2}$ following \citet{C18}, and we assume a linear relationship between $A_{\rm K}$ and $\Sigma_{\rm gas}$.
    \label{fig:L10}
    }
\end{figure}

We next consider the method for detecting thresholds, and the associated evidence for a threshold, presented by \citet[hereafter L10]{L10}. L10 survey 11 nearby clouds and produce IR extinction maps. They count the number of YSOs within contours drawn on clouds and show that the relation $N_{\rm YSO} \approx M_{\rm gas}^{0.96 \pm 0.11}$ is satisfied for these clouds when $M_{\rm gas}$ is taken to be the total mass of gas above an extinction threshold of 0.8 $A_\mathrm{K \ mag}$. L10 find that the cloud-to-cloud scatter in the ratio $M_{\rm gas} / N_{\rm YSO}$ (or equivalently $N_{\rm YSO} / M_{\rm gas}$) between the 11 local clouds surveyed is minimised if one counts mass and YSOs within a contour corresponding to an extinction $A_\mathrm{K} = 0.8 \ \mathrm{mag}$. They argue based on this that gas at extinction above $A_\mathrm{K} = 0.8 \ \mathrm{mag}$ constitutes the active material for star formation, so that $A_\mathrm{K} = 0.8 \ \mathrm{mag}$ constitutes a threshold. 

In \autoref{fig:L10}, we show the $N_{\rm YSO} / M_{\rm gas}$ as a function of contour level $\Sigma_{\rm gas}$ for the lower resolution (Res1) mock observations from simulations alongside the data from L10; we use the Res1 rather than Res2 case because this is more typical of NIR extinction methods. We find that the mock data from the simulations lie in close agreement with the observed data, and that the scatter between the results for the three projection axes for a single simulation is comparable to the observed cloud-to-cloud scatter. However, examining \autoref{fig:L10} it is clear that the simulation points do not display any feature through which one could recover the physical volume density threshold that we have shown is present. The simulations and observations both do show a rise in $N_{\rm YSO}/M_{\rm gas}$ at the highest contour levels. This could plausibly be attributed to a rise in star formation efficiency at the highest column densities, but we note that these features are (1) at extinction values far higher than the purported threshold of $0.8$ mag in K-band luminosity for which L10 argue, and (2) as a result of low resolution are at column densities that are significantly below the true threshold discussed in \autoref{sec:eff-rho-gas} and \autoref{sec:eff-sigma-gas}.

\section{Summary and Conclusions}
\label{sec:Conc}

In this paper we use simulations of star formation including a wide range of physical processes to probe the existence of star formation thresholds, and we investigate whether it is possible to measure them using current observational techniques.
We show that the simulations we analyse do possess a true local volume density threshold (different from the sink particle threshold) of $n_{\rm th} \sim 10^6 - 10^7$ cm$^{-3}$; gas above this threshold appears to form stars efficiently, without substantial opposition from turbulence, magnetic fields or outflow feedback. The evidence for this is a drastic change in the star formation efficiency per free-fall time $\epsilon_{\rm ff}$ across the threshold, a feature that naturally defines a true star formation threshold in the simulations. 

We then investigate whether it is possible to recover this information using only the 2D, static information that is available in observations. We create projection maps from the simulations, which we then blur to resolutions typical of mid-infrared dust emission or near-infrared dust extinction studies. We find that when we estimate $\epsilon_{\rm ff}$ from un-smoothed projection maps, we can still recover the characteristic rising $\epsilon_{\rm ff}$ feature at column density values of $\Sigma_{\rm gas, th} \sim 2 \times 10^3 \ \mathrm{M}_\odot \ \mathrm{pc}^{-2}$, and we propose a method to estimate an associated volume density that recovers the true, 3D threshold volume density reasonably well. For the mock observations at realistic resolution, we find that plots of projected $\epsilon_{\rm ff}$ remain a reliable indicator of the presence of a threshold, as shown by a rising $\epsilon_{\rm ff}$ feature, but that the resolution currently typical of such observations renders recovery of the true column or volume density a challenge. In contrast, methods based on comparing $\Sigma_{\rm SFR}$ and $\Sigma_{\rm gas}$, or on measuring the ratio of young stellar object number to gas mass, provide no evidence for a volume or column density threshold for star formation in our simulations, despite the fact that one exists. We therefore conclude that these methods are not reliable indicators for the presence of a threshold for star formation.

\section*{Acknowledgements}

We would like to thank the anonymous referee whose useful comments have helped improve the manuscript. MRK and CF both acknowledge support from the Australian Research Council's (ARC) Discovery Projects and Future Fellowship funding schemes, awards DP160100695 (MRK), DP170100603 (CF), FT180100375 (MRK), and FT180100495 (CF), and from the Australia-Germany Joint Research Cooperation Scheme (UA-DAAD). AJC's work was performed under the auspices of the U.S. Department of Energy by Lawrence Livermore National Laboratory under Contract DE-AC52-07NA27344. The simulations used in this work were made possible by grants of high-performance computing resources from the following: the National Center of Supercomputing Application through grant TGMCA00N020, under the Extreme Science and Engineering Discovery Environment, which is supported by National Science Foundation grant number OCI-1053575; the NASA High-End Computing Program through the NASA Advanced Supercomputing (NAS) Division at Ames Research Center; the Leibniz Rechenzentrum and the Gauss Centre for Supercomputing (grants~pr32lo, pr48pi and GCS Large-scale project~10391); the Partnership for Advanced Computing in Europe (PRACE grant pr89mu); the National Computational Infrastructure, which is supported by the Australian Government (grants ek9 and jh2); the Pawsey Supercomputing Centre with funding from the Australian Government and the Government of Western Australia. The simulation software FLASH was in part developed by the DOE-supported Flash Center for Computational Science at the University of Chicago.   LLNL-JRNL-766677

\bibliographystyle{mnras}
\bibliography{references.bib}

\begin{thebibliography}{}
\makeatletter
\relax
\def\mn@urlcharsother{\let\do\@makeother \do\$\do\&\do\#\do\^\do\_\do\%\do\~}
\def\mn@doi{\begingroup\mn@urlcharsother \@ifnextchar [ {\mn@doi@}
  {\mn@doi@[]}}
\def\mn@doi@[#1]#2{\def\@tempa{#1}\ifx\@tempa\@empty \href
  {http://dx.doi.org/#2} {doi:#2}\else \href {http://dx.doi.org/#2} {#1}\fi
  \endgroup}
\def\mn@eprint#1#2{\mn@eprint@#1:#2::\@nil}
\def\mn@eprint@arXiv#1{\href {http://arxiv.org/abs/#1} {{\tt arXiv:#1}}}
\def\mn@eprint@dblp#1{\href {http://dblp.uni-trier.de/rec/bibtex/#1.xml}
  {dblp:#1}}
\def\mn@eprint@#1:#2:#3:#4\@nil{\def\@tempa {#1}\def\@tempb {#2}\def\@tempc
  {#3}\ifx \@tempc \@empty \let \@tempc \@tempb \let \@tempb \@tempa \fi \ifx
  \@tempb \@empty \def\@tempb {arXiv}\fi \@ifundefined
  {mn@eprint@\@tempb}{\@tempb:\@tempc}{\expandafter \expandafter \csname
  mn@eprint@\@tempb\endcsname \expandafter{\@tempc}}}

\bibitem[\protect\citeauthoryear{{Andr{\'e}}}{{Andr{\'e}}}{2015}]{Andre15}
{Andr{\'e}} P.,  2015, \mn@doi [Highlights of Astronomy]
  {10.1017/S1743921314004633}, \href
  {https://ui.adsabs.harvard.edu/#abs/2015HiA....16...31A} {16, 31}

\bibitem[\protect\citeauthoryear{{Andr{\'e}}}{{Andr{\'e}}}{2017}]{Andre2017}
{Andr{\'e}} P.,  2017, \mn@doi [Comptes Rendus Geoscience]
  {https://doi.org/10.1016/j.crte.2017.07.002}, 349, 187

\bibitem[\protect\citeauthoryear{{Andr{\'e}} et~al.,}{{Andr{\'e}}
  et~al.}{2010}]{HerschelRes2010}
{Andr{\'e}} P.,  et~al., 2010, \mn@doi [\aap] {10.1051/0004-6361/201014666},
  \href {http://cdsads.u-strasbg.fr/abs/2010A%26A...518L.102A} {518, L102}

\bibitem[\protect\citeauthoryear{{Andr{\'e}}, {Di Francesco}, {Ward-Thompson},
  {Inutsuka}, {Pudritz}  \& {Pineda}}{{Andr{\'e}} et~al.}{2014}]{A14}
{Andr{\'e}} P.,  {Di Francesco} J.,  {Ward-Thompson} D.,  {Inutsuka} S.~I.,
  {Pudritz} R.~E.,   {Pineda} J.~E.,  2014, in {Beuther} H.,  {Klessen} R.~S.,
  {Dullemond} C.~P.,   {Henning} T.,  eds, Protostars and Planets VI. p.~27
  (\mn@eprint {arXiv} {1312.6232}),
  \mn@doi{10.2458/azu_uapress_9780816531240-ch002}

\bibitem[\protect\citeauthoryear{{Basu} \& {Ciolek}}{{Basu} \&
  {Ciolek}}{2004}]{Basu04magnetic}
{Basu} S.,  {Ciolek} G.~E.,  2004, \mn@doi [\apj] {10.1086/421464}, \href
  {https://ui.adsabs.harvard.edu/#abs/2004ApJ...607L..39B} {607, L39}

\bibitem[\protect\citeauthoryear{{Berger} \& {Colella}}{{Berger} \&
  {Colella}}{1989}]{Berger1989}
{Berger} M.~J.,  {Colella} P.,  1989, \mn@doi [Journal of Computational
  Physics] {10.1016/0021-9991(89)90035-1}, \href
  {http://adsabs.harvard.edu/abs/1989JCoPh..82...64B} {82, 64}

\bibitem[\protect\citeauthoryear{{Bigiel} et~al.,}{{Bigiel}
  et~al.}{2016}]{Bigiel16a}
{Bigiel} F.,  et~al., 2016, \apjl, 822, L26

\bibitem[\protect\citeauthoryear{{Blanc}, {Heiderman}, {Gebhardt}, {Evans}  \&
  {Adams}}{{Blanc} et~al.}{2009}]{Blanc2009}
{Blanc} G.~A.,  {Heiderman} A.,  {Gebhardt} K.,  {Evans} Neal~J. I.,   {Adams}
  J.,  2009, \mn@doi [\apj] {10.1088/0004-637X/704/1/842}, \href
  {https://ui.adsabs.harvard.edu/#abs/2009ApJ...704..842B} {704, 842}

\bibitem[\protect\citeauthoryear{{Burkert} \& {Hartmann}}{{Burkert} \&
  {Hartmann}}{2013}]{BH13}
{Burkert} A.,  {Hartmann} L.,  2013, \mn@doi [\apj]
  {10.1088/0004-637X/773/1/48}, \href
  {https://ui.adsabs.harvard.edu/#abs/2013ApJ...773...48B} {773, 48}

\bibitem[\protect\citeauthoryear{{Clark} \& {Glover}}{{Clark} \&
  {Glover}}{2014}]{CG14}
{Clark} P.~C.,  {Glover} S. C.~O.,  2014, \mn@doi [\mnras]
  {10.1093/mnras/stu1589}, \href
  {https://ui.adsabs.harvard.edu/#abs/2014MNRAS.444.2396C} {444, 2396}

\bibitem[\protect\citeauthoryear{{Cunningham}, {Klein}, {Krumholz}  \&
  {McKee}}{{Cunningham} et~al.}{2011}]{Cunningham2011}
{Cunningham} A.~J.,  {Klein} R.~I.,  {Krumholz} M.~R.,   {McKee} C.~F.,  2011,
  \mn@doi [\apj] {10.1088/0004-637X/740/2/107}, \href
  {https://ui.adsabs.harvard.edu/#abs/2011ApJ...740..107C} {740, 107}

\bibitem[\protect\citeauthoryear{{Cunningham}, {Krumholz}, {McKee}  \&
  {Klein}}{{Cunningham} et~al.}{2018}]{C18}
{Cunningham} A.~J.,  {Krumholz} M.~R.,  {McKee} C.~F.,   {Klein} R.~I.,  2018,
  \mn@doi [\mnras] {10.1093/mnras/sty154}, \href
  {https://ui.adsabs.harvard.edu/#abs/2018MNRAS.476..771C} {476, 771}

\bibitem[\protect\citeauthoryear{Dubey et~al.,}{Dubey et~al.}{2008}]{Dubey2008}
Dubey A.,  et~al., 2008, Astronomical Society of the Pacific Conference Series,
  385, 145

\bibitem[\protect\citeauthoryear{{Elmegreen}}{{Elmegreen}}{2018}]{Elmegreen18a}
{Elmegreen} B.~G.,  2018, \mn@doi [\apj] {10.3847/1538-4357/aaa770}, \href
  {https://ui.adsabs.harvard.edu/abs/2018ApJ...854...16E} {854, 16}

\bibitem[\protect\citeauthoryear{{Evans}, {Heiderman}  \&
  {Vutisalchavakul}}{{Evans} et~al.}{2014}]{Evans14a}
{Evans} II N.~J.,  {Heiderman} A.,   {Vutisalchavakul} N.,  2014, \apj, 782,
  114

\bibitem[\protect\citeauthoryear{{Federrath}}{{Federrath}}{2013a}]{F13}
{Federrath} C.,  2013a, \mn@doi [\mnras] {10.1093/mnras/stt1644}, \href
  {https://ui.adsabs.harvard.edu/#abs/2013MNRAS.436.1245F} {436, 1245}

\bibitem[\protect\citeauthoryear{{Federrath}}{{Federrath}}{2013b}]{Federrath2013}
{Federrath} C.,  2013b, \mn@doi [\mnras] {10.1093/mnras/stt1799}, \href
  {https://ui.adsabs.harvard.edu/\#abs/2013MNRAS.436.3167F} {436, 3167}

\bibitem[\protect\citeauthoryear{{Federrath}}{{Federrath}}{2015}]{F15}
{Federrath} C.,  2015, \mn@doi [\mnras] {10.1093/mnras/stv941}, \href
  {http://adsabs.harvard.edu/abs/2015MNRAS.450.4035F} {450, 4035}

\bibitem[\protect\citeauthoryear{{Federrath} \& {Klessen}}{{Federrath} \&
  {Klessen}}{2012}]{FK12}
{Federrath} C.,  {Klessen} R.~S.,  2012, \mn@doi [\apj]
  {10.1088/0004-637X/761/2/156}, \href
  {http://adsabs.harvard.edu/abs/2012ApJ...761..156F} {761, 156}

\bibitem[\protect\citeauthoryear{{Federrath} \& {Klessen}}{{Federrath} \&
  {Klessen}}{2013}]{FK13}
{Federrath} C.,  {Klessen} R.~S.,  2013, \mn@doi [\apj]
  {10.1088/0004-637X/763/1/51}, \href
  {https://ui.adsabs.harvard.edu/#abs/2013ApJ...763...51F} {763, 51}

\bibitem[\protect\citeauthoryear{{Federrath}, {Roman-Duval}, {Klessen},
  {Schmidt}  \& {Mac Low}}{{Federrath} et~al.}{2010a}]{Federrath2010}
{Federrath} C.,  {Roman-Duval} J.,  {Klessen} R.~S.,  {Schmidt} W.,   {Mac Low}
  M.~M.,  2010a, \mn@doi [\aap] {10.1051/0004-6361/200912437}, \href
  {https://ui.adsabs.harvard.edu/#abs/2010A&A...512A..81F} {512, A81}

\bibitem[\protect\citeauthoryear{{Federrath}, {Banerjee}, {Clark}  \&
  {Klessen}}{{Federrath} et~al.}{2010b}]{FederrathEtAl2010}
{Federrath} C.,  {Banerjee} R.,  {Clark} P.~C.,   {Klessen} R.~S.,  2010b,
  \mn@doi [\apj] {10.1088/0004-637X/713/1/269}, \href
  {https://ui.adsabs.harvard.edu/#abs/2010ApJ...713..269F} {713, 269}

\bibitem[\protect\citeauthoryear{{Federrath}, {Schr{\"o}n}, {Banerjee}  \&
  {Klessen}}{{Federrath} et~al.}{2014}]{FederrathEtAl2014}
{Federrath} C.,  {Schr{\"o}n} M.,  {Banerjee} R.,   {Klessen} R.~S.,  2014,
  \mn@doi [\apj] {10.1088/0004-637X/790/2/128}, \href
  {https://ui.adsabs.harvard.edu/#abs/2014ApJ...790..128F} {790, 128}

\bibitem[\protect\citeauthoryear{{Federrath}, {Krumholz}  \&
  {Hopkins}}{{Federrath} et~al.}{2017}]{FKH17}
{Federrath} C.,  {Krumholz} M.,   {Hopkins} P.~F.,  2017, in Journal of Physics
  Conference Series. p. 012007, \mn@doi{10.1088/1742-6596/837/1/012007}

\bibitem[\protect\citeauthoryear{{Fryxell} et~al.,}{{Fryxell}
  et~al.}{2000}]{Fryxell2000}
{Fryxell} B.,  et~al., 2000, \mn@doi [\apjs] {10.1086/317361}, \href
  {http://adsabs.harvard.edu/abs/2000ApJS..131..273F} {131, 273}

\bibitem[\protect\citeauthoryear{{Garc{\'{\i}}a-Burillo}, {Usero},
  {Alonso-Herrero}, {Graci{\'a}-Carpio}, {Pereira-Santaella}, {Colina},
  {Planesas}  \& {Arribas}}{{Garc{\'{\i}}a-Burillo}
  et~al.}{2012}]{Garcia-Burillo12a}
{Garc{\'{\i}}a-Burillo} S.,  {Usero} A.,  {Alonso-Herrero} A.,
  {Graci{\'a}-Carpio} J.,  {Pereira-Santaella} M.,  {Colina} L.,  {Planesas}
  P.,   {Arribas} S.,  2012, \aap, 539, A8

\bibitem[\protect\citeauthoryear{{Goldsmith}, {Heyer}, {Narayanan}, {Snell},
  {Li}  \& {Brunt}}{{Goldsmith} et~al.}{2008}]{Goldsmith08filament}
{Goldsmith} P.~F.,  {Heyer} M.,  {Narayanan} G.,  {Snell} R.,  {Li} D.,
  {Brunt} C.,  2008, \mn@doi [\apj] {10.1086/587166}, \href
  {https://ui.adsabs.harvard.edu/#abs/2008ApJ...680..428G} {680, 428}

\bibitem[\protect\citeauthoryear{{Goodman}, {Pineda}  \& {Schnee}}{{Goodman}
  et~al.}{2009}]{Goodman2009}
{Goodman} A.~A.,  {Pineda} J.~E.,   {Schnee} S.~L.,  2009, \mn@doi [\apj]
  {10.1088/0004-637X/692/1/91}, \href
  {https://ui.adsabs.harvard.edu/\#abs/2009ApJ...692...91G} {692, 91}

\bibitem[\protect\citeauthoryear{{Gutermuth}, {Pipher}, {Megeath}, {Myers},
  {Allen}  \& {Allen}}{{Gutermuth} et~al.}{2011}]{Gutermuth11a}
{Gutermuth} R.~A.,  {Pipher} J.~L.,  {Megeath} S.~T.,  {Myers} P.~C.,  {Allen}
  L.~E.,   {Allen} T.~S.,  2011, \mn@doi [\apj] {10.1088/0004-637X/739/2/84},
  \href {http://adsabs.harvard.edu/abs/2011ApJ...739...84G} {739, 84}

\bibitem[\protect\citeauthoryear{{Hatchell}, {Richer}, {Fuller}, {Qualtrough},
  {Ladd}  \& {Chandler}}{{Hatchell} et~al.}{2005}]{Hatchell05}
{Hatchell} J.,  {Richer} J.~S.,  {Fuller} G.~A.,  {Qualtrough} C.~J.,  {Ladd}
  E.~F.,   {Chandler} C.~J.,  2005, \mn@doi [\aap]
  {10.1051/0004-6361:20041836}, \href
  {https://ui.adsabs.harvard.edu/#abs/2005A&A...440..151H} {440, 151}

\bibitem[\protect\citeauthoryear{{Heiderman}, {Evans}, {Allen}, {Huard}  \&
  {Heyer}}{{Heiderman} et~al.}{2010}]{H10}
{Heiderman} A.,  {Evans} II N.~J.,  {Allen} L.~E.,  {Huard} T.,   {Heyer} M.,
  2010, \mn@doi [\apj] {10.1088/0004-637X/723/2/1019}, \href
  {http://adsabs.harvard.edu/abs/2010ApJ...723.1019H} {723, 1019}

\bibitem[\protect\citeauthoryear{{Hennebelle} \& {Chabrier}}{{Hennebelle} \&
  {Chabrier}}{2011}]{HC11}
{Hennebelle} P.,  {Chabrier} G.,  2011, \mn@doi [\apjl]
  {10.1088/2041-8205/743/2/L29}, \href
  {http://adsabs.harvard.edu/abs/2011ApJ...743L..29H} {743, L29}

\bibitem[\protect\citeauthoryear{{Heyer}, {Gutermuth}, {Urquhart}, {Csengeri},
  {Wienen}, {Leurini}, {Menten}  \& {Wyrowski}}{{Heyer}
  et~al.}{2016}]{Heyer16a}
{Heyer} M.,  {Gutermuth} R.,  {Urquhart} J.~S.,  {Csengeri} T.,  {Wienen} M.,
  {Leurini} S.,  {Menten} K.,   {Wyrowski} F.,  2016, \aap, 588, A29

\bibitem[\protect\citeauthoryear{{Hopkins}}{{Hopkins}}{2012}]{Hopkins2012a}
{Hopkins} P.~F.,  2012, \mn@doi [\mnras] {10.1111/j.1365-2966.2012.20730.x},
  \href {https://ui.adsabs.harvard.edu/#abs/2012MNRAS.423.2016H} {423, 2016}

\bibitem[\protect\citeauthoryear{{Hopkins}}{{Hopkins}}{2013}]{Hopkins2013a}
{Hopkins} P.~F.,  2013, \mn@doi [\mnras] {10.1093/mnras/sts704}, \href
  {https://ui.adsabs.harvard.edu/#abs/2013MNRAS.430.1653H} {430, 1653}

\bibitem[\protect\citeauthoryear{{Johnstone}, {Di Francesco}  \&
  {Kirk}}{{Johnstone} et~al.}{2004}]{Johnstone04}
{Johnstone} D.,  {Di Francesco} J.,   {Kirk} H.,  2004, \mn@doi [\apj]
  {10.1086/423737}, \href
  {https://ui.adsabs.harvard.edu/#abs/2004ApJ...611L..45J} {611, L45}

\bibitem[\protect\citeauthoryear{{Juvela} \& {Montillaud}}{{Juvela} \&
  {Montillaud}}{2016}]{NICESTRes2016}
{Juvela} M.,  {Montillaud} J.,  2016, \mn@doi [\aap]
  {10.1051/0004-6361/201425112}, \href
  {http://adsabs.harvard.edu/abs/2016A%26A...585A..38J} {585, A38}

\bibitem[\protect\citeauthoryear{{Kennicutt} Robert~C. et~al.,}{{Kennicutt}
  et~al.}{2007}]{Kennicutt2007}
{Kennicutt} Robert~C. J.,  et~al., 2007, \mn@doi [\apj] {10.1086/522300}, \href
  {https://ui.adsabs.harvard.edu/#abs/2007ApJ...671..333K} {671, 333}

\bibitem[\protect\citeauthoryear{{Klein}, {Fisher}, {McKee}  \&
  {Truelove}}{{Klein} et~al.}{1999}]{Klein1999}
{Klein} R.~I.,  {Fisher} R.~T.,  {McKee} C.~F.,   {Truelove} J.~K.,  1999, in
  {Miyama} S.~M.,  {Tomisaka} K.,   {Hanawa} T.,  eds,  Vol. 240, Numerical
  Astrophysics. p.~131 (\mn@eprint {arXiv} {astro-ph/9806330}),
  \mn@doi{10.1007/978-94-011-4780-4_44}

\bibitem[\protect\citeauthoryear{{K{\"o}nyves} et~al.,}{{K{\"o}nyves}
  et~al.}{2015}]{Konyves15}
{K{\"o}nyves} V.,  et~al., 2015, \mn@doi [\aap] {10.1051/0004-6361/201525861},
  \href {https://ui.adsabs.harvard.edu/#abs/2015A&A...584A..91K} {584, A91}

\bibitem[\protect\citeauthoryear{{Krumholz}}{{Krumholz}}{2014}]{Krumholz14}
{Krumholz} M.~R.,  2014, \mn@doi [\physrep] {10.1016/j.physrep.2014.02.001},
  \href {https://ui.adsabs.harvard.edu/#abs/2014PhR...539...49K} {539, 49}

\bibitem[\protect\citeauthoryear{{Krumholz} \& {McKee}}{{Krumholz} \&
  {McKee}}{2005}]{KM05}
{Krumholz} M.~R.,  {McKee} C.~F.,  2005, \mn@doi [\apj] {10.1086/431734}, \href
  {http://adsabs.harvard.edu/abs/2005ApJ...630..250K} {630, 250}

\bibitem[\protect\citeauthoryear{{Krumholz} \& {Tan}}{{Krumholz} \&
  {Tan}}{2007}]{KrumholzTan2007}
{Krumholz} M.~R.,  {Tan} J.~C.,  2007, \mn@doi [\apj] {10.1086/509101}, \href
  {http://adsabs.harvard.edu/abs/2007ApJ...654..304K} {654, 304}

\bibitem[\protect\citeauthoryear{{Krumholz} \& {Thompson}}{{Krumholz} \&
  {Thompson}}{2007}]{KrumholzThompson07}
{Krumholz} M.~R.,  {Thompson} T.~A.,  2007, \mn@doi [\apj] {10.1086/521642},
  \href {http://adsabs.harvard.edu/abs/2007ApJ...669..289K} {669, 289}

\bibitem[\protect\citeauthoryear{{Krumholz}, {McKee}  \& {Klein}}{{Krumholz}
  et~al.}{2004}]{Krumholz04}
{Krumholz} M.~R.,  {McKee} C.~F.,   {Klein} R.~I.,  2004, \mn@doi [\apj]
  {10.1086/421935}, \href {http://adsabs.harvard.edu/abs/2004ApJ...611..399K}
  {611, 399}

\bibitem[\protect\citeauthoryear{{Krumholz}, {Klein}, {McKee}  \&
  {Bolstad}}{{Krumholz} et~al.}{2007}]{Krumholzetal2007}
{Krumholz} M.~R.,  {Klein} R.~I.,  {McKee} C.~F.,   {Bolstad} J.,  2007,
  \mn@doi [\apj] {10.1086/520791}, \href
  {https://ui.adsabs.harvard.edu/#abs/2007ApJ...667..626K} {667, 626}

\bibitem[\protect\citeauthoryear{{Krumholz}, {Dekel}  \& {McKee}}{{Krumholz}
  et~al.}{2012}]{KDM12}
{Krumholz} M.~R.,  {Dekel} A.,   {McKee} C.~F.,  2012, \mn@doi [\apj]
  {10.1088/0004-637X/745/1/69}, \href
  {http://adsabs.harvard.edu/abs/2012ApJ...745...69K} {745, 69}

\bibitem[\protect\citeauthoryear{{Krumholz}, {McKee}  \&
  {Bland-Hawthorn}}{{Krumholz} et~al.}{2019}]{Krumholz18a}
{Krumholz} M.~R.,  {McKee} C.~F.,   {Bland-Hawthorn} J.,  2019, \araa, pp in
  press, arXiv:1812.01615

\bibitem[\protect\citeauthoryear{{Lada}, {Lombardi}  \& {Alves}}{{Lada}
  et~al.}{2010}]{L10}
{Lada} C.~J.,  {Lombardi} M.,   {Alves} J.~F.,  2010, \mn@doi [\apj]
  {10.1088/0004-637X/724/1/687}, \href
  {http://adsabs.harvard.edu/abs/2010ApJ...724..687L} {724, 687}

\bibitem[\protect\citeauthoryear{{Lada}, {Forbrich}, {Lombardi}  \&
  {Alves}}{{Lada} et~al.}{2012}]{Lada12}
{Lada} C.~J.,  {Forbrich} J.,  {Lombardi} M.,   {Alves} J.~F.,  2012, \mn@doi
  [\apj] {10.1088/0004-637X/745/2/190}, \href
  {https://ui.adsabs.harvard.edu/#abs/2012ApJ...745..190L} {745, 190}

\bibitem[\protect\citeauthoryear{{Leroy} et~al.,}{{Leroy}
  et~al.}{2017}]{Leroy17a}
{Leroy} A.~K.,  et~al., 2017, \apj, 846, 71

\bibitem[\protect\citeauthoryear{Li, Martin, Klein  \& McKee}{Li
  et~al.}{2012}]{Lietal2012}
Li P.~S.,  Martin D.~F.,  Klein R.~I.,   McKee C.~F.,  2012, The Astrophysical
  Journal, 745, 139

\bibitem[\protect\citeauthoryear{{Mac Low}}{{Mac Low}}{1999}]{MacLow1999}
{Mac Low} M.-M.,  1999, \mn@doi [\apj] {10.1086/307784}, \href
  {https://ui.adsabs.harvard.edu/#abs/1999ApJ...524..169M} {524, 169}

\bibitem[\protect\citeauthoryear{{McKee}}{{McKee}}{1989}]{McKee89a}
{McKee} C.~F.,  1989, \apj, 345, 782

\bibitem[\protect\citeauthoryear{{Mouschovias} \& {Spitzer}}{{Mouschovias} \&
  {Spitzer}}{1976}]{Mouschovias1976}
{Mouschovias} T.~C.,  {Spitzer} Jr. L.,  1976, \mn@doi [\apj] {10.1086/154835},
  \href {http://adsabs.harvard.edu/abs/1976ApJ...210..326M} {210, 326}

\bibitem[\protect\citeauthoryear{{Offner}, {Klein}, {McKee}  \&
  {Krumholz}}{{Offner} et~al.}{2009}]{Offner2009}
{Offner} S. S.~R.,  {Klein} R.~I.,  {McKee} C.~F.,   {Krumholz} M.~R.,  2009,
  \mn@doi [\apj] {10.1088/0004-637X/703/1/131}, \href
  {https://ui.adsabs.harvard.edu/#abs/2009ApJ...703..131O} {703, 131}

\bibitem[\protect\citeauthoryear{{Onishi}, {Mizuno}, {Kawamura}, {Ogawa}  \&
  {Fukui}}{{Onishi} et~al.}{1998}]{Onishi98}
{Onishi} T.,  {Mizuno} A.,  {Kawamura} A.,  {Ogawa} H.,   {Fukui} Y.,  1998,
  \mn@doi [\apj] {10.1086/305867}, \href
  {https://ui.adsabs.harvard.edu/#abs/1998ApJ...502..296O} {502, 296}

\bibitem[\protect\citeauthoryear{{Onus}, {Krumholz}  \& {Federrath}}{{Onus}
  et~al.}{2018}]{Onus18}
{Onus} A.,  {Krumholz} M.~R.,   {Federrath} C.,  2018, \mn@doi [\mnras]
  {10.1093/mnras/sty1662}, \href
  {https://ui.adsabs.harvard.edu/#abs/2018MNRAS.479.1702O} {479, 1702}

\bibitem[\protect\citeauthoryear{Padoan \& Nordlund}{Padoan \&
  Nordlund}{2011}]{PN11}
Padoan P.,  Nordlund A.,  2011, The Astrophysical Journal, 730, 40

\bibitem[\protect\citeauthoryear{{Salim}, {Federrath}  \& {Kewley}}{{Salim}
  et~al.}{2015}]{SFK15}
{Salim} D.~M.,  {Federrath} C.,   {Kewley} L.~J.,  2015, \mn@doi [\apj]
  {10.1088/2041-8205/806/2/L36}, \href
  {https://ui.adsabs.harvard.edu/#abs/2015ApJ...806L..36S} {806, L36}

\bibitem[\protect\citeauthoryear{{Sharda}, {Federrath}, {da Cunha}, {Swinbank}
  \& {Dye}}{{Sharda} et~al.}{2018}]{Sharda18}
{Sharda} P.,  {Federrath} C.,  {da Cunha} E.,  {Swinbank} A.~M.,   {Dye} S.,
  2018, \mn@doi [\mnras] {10.1093/mnras/sty886}, \href
  {https://ui.adsabs.harvard.edu/#abs/2018MNRAS.477.4380S} {477, 4380}

\bibitem[\protect\citeauthoryear{{Shimajiri} et~al.,}{{Shimajiri}
  et~al.}{2017}]{Shimajiri17a}
{Shimajiri} Y.,  et~al., 2017, \aap, 604, A74

\bibitem[\protect\citeauthoryear{{Shu}, {Adams}  \& {Lizano}}{{Shu}
  et~al.}{1987}]{Shu1987}
{Shu} F.~H.,  {Adams} F.~C.,   {Lizano} S.,  1987, \mn@doi [Annual Review of
  Astronomy and Astrophysics] {10.1146/annurev.aa.25.090187.000323}, \href
  {https://ui.adsabs.harvard.edu/#abs/1987ARA&A..25...23S} {25, 23}

\bibitem[\protect\citeauthoryear{{Strai{\v{z}}ys}, {{\v{C}}ernis}  \&
  {Barta{\v{s}}i{\={u}}t{\.{e}}}}{{Strai{\v{z}}ys}
  et~al.}{2003}]{Aquiladist2003}
{Strai{\v{z}}ys} V.,  {{\v{C}}ernis} K.,   {Barta{\v{s}}i{\={u}}t{\.{e}}} S.,
  2003, \mn@doi [\aap] {10.1051/0004-6361:20030599}, \href
  {https://ui.adsabs.harvard.edu/#abs/2003A&A...405..585S} {405, 585}

\bibitem[\protect\citeauthoryear{{Torres}, {Loinard}, {Mioduszewski}  \&
  {Rodr{\'{\i}}guez}}{{Torres} et~al.}{2007}]{Taurus}
{Torres} R.~M.,  {Loinard} L.,  {Mioduszewski} A.~J.,   {Rodr{\'{\i}}guez}
  L.~F.,  2007, \mn@doi [\apj] {10.1086/522924}, \href
  {http://adsabs.harvard.edu/abs/2007ApJ...671.1813T} {671, 1813}

\bibitem[\protect\citeauthoryear{{Truelove}, {Klein}, {McKee}, {Holliman},
  {Howell}, {Greenough}  \& {Woods}}{{Truelove} et~al.}{1998}]{Truelove1998}
{Truelove} J.~K.,  {Klein} R.~I.,  {McKee} C.~F.,  {Holliman} II J.~H.,
  {Howell} L.~H.,  {Greenough} J.~A.,   {Woods} D.~T.,  1998, \mn@doi [\apj]
  {10.1086/305329}, \href {http://adsabs.harvard.edu/abs/1998ApJ...495..821T}
  {495, 821}

\bibitem[\protect\citeauthoryear{{Usero} et~al.,}{{Usero}
  et~al.}{2015}]{Usero15a}
{Usero} A.,  et~al., 2015, \aj, 150, 115

\bibitem[\protect\citeauthoryear{{Vutisalchavakul}, {Evans}  \&
  {Heyer}}{{Vutisalchavakul} et~al.}{2016}]{Vutisalchavakul16a}
{Vutisalchavakul} N.,  {Evans} II N.~J.,   {Heyer} M.,  2016, \apj, 831, 73

\bibitem[\protect\citeauthoryear{{Wong} \& {Blitz}}{{Wong} \&
  {Blitz}}{2002}]{WongBlitz2002}
{Wong} T.,  {Blitz} L.,  2002, \mn@doi [\apj] {10.1086/339287}, \href
  {https://ui.adsabs.harvard.edu/#abs/2002ApJ...569..157W} {569, 157}

\makeatother
\end{thebibliography}

\bsp	
\label{lastpage}
\end{document}